\documentstyle[preprint,prl,aps]{revtex}
\begin{document}
\tightenlines
\draft
\title{ General relativistic gravitational field of a rigidly rotating
disk of dust: Solution in terms of ultraelliptic functions}
\author{ G. Neugebauer and R. Meinel}
\address{Max--Planck--Gesellschaft, 
Arbeitsgruppe Gravitationstheorie \\ an der Universit\"at 
Jena, Max-Wien-Platz 1, D-07743 Jena, Germany}
\maketitle
\begin{abstract}
In a recent paper we presented analytic expressions for
the axis potential,
the disk metric, and the surface mass density of the global solution
to Einstein's field equations describing a
rigidly rotating disk of dust.
Here we add the complete solution in terms of ultraelliptic
functions and quadratures.
\end{abstract}
\paragraph{Introduction.}
An infinitesimally thin disk of dust rotating uniformly around 
its symmetry axis is the simplest model of a rotating 
self--gravitating body.  Within Newton's theory of gravitation this
model is described by the Maclaurin solution of the Laplace equation.
In the present letter we show that the model of a 
uniformly rotating disk of dust
allows for an explicit analytic solution in Einstein's theory of
gravitation  as well.
This seems to be the first rigorous global solution of the rotating body
problem (for perfect fluids with $p=0$) in general relativity.

An approximate solution of the problem was given by Bardeen and 
Wagoner \cite{BW1}, \cite{BW2}. By applying the Inverse (Scattering)
Method to the related boundary value problem of the Ernst
equation we found the solution in terms of two linear integral
equations, a `small' and a `big' one \cite{NM}. The analytic solution
to the small integral equation in terms of the Weierstrass function
led to explicit expressions for the Ernst potential on the symmetry
axis, the disk metric, and the surface mass density \cite{NM2}.  
In the following we will show that the complete solution of the
problem may be represented, up to quadratures,
in terms of {\it ultraelliptic} functions.   

\paragraph{The metric.}
The line element reads in Weyl--Lewis--Papapetrou--coordinates:
\begin{equation}
ds^2 = e^{-2U}[e^{2k}(d\rho^2+d\zeta^2)+\rho^2d\varphi^2]-
e^{2U}(dt+ad\varphi)^2.
\end{equation}
(We use units where Newton's 
gravitational constant $G$ as well as
the velocity of light $c$ are equal to 1.)
The metric functions $e^{2U}$, $e^{2k}$,
and $a/\rho_0$ depend  uniquely on the normalized coordinates
$\rho/\rho_0$, $\zeta/\rho_0$,
and the parameter 
\begin{equation}
\mu=2\Omega^2\rho_0^2e^{-2V_0},
\label{mu}
\end{equation}  
where $\Omega$, $\rho_0$ and $V_0$ are the angular velocity, 
the coordinate radius,
and the `surface potential' 
$V_o(\mu) \equiv U(\rho=0,\zeta=0,\mu$), 
respectively. The disk is characterized by $\zeta=0$, 
$0\le\rho\le \rho_0$. 

\paragraph{The Ernst potential.}
The (complex) Ernst potential is defined by 
\begin{equation}
f = e^{2U} + {\rm i} b,
\end{equation}
where the imaginary part $b$ is related to the 
metric function $a$ according to
\begin{equation}
a_{,\rho} =  \rho\,e^{-4U}b_{,\zeta}; 
\quad  a_{,\zeta} = - \rho\,e^{-4U}b_{,\rho}
\end{equation}
i.~e., $a$ may be calculated from a given Ernst potential (and its
derivatives) by a quadrature. The same holds for the metric
function $e^{2k}$. Hence, an axisymmetric stationary solution to the 
vacuum Einstein equations is sufficiently characterized by its Ernst
potential. Note that our global solution is completely described by the vacuum
solution outside the disk including the boundary data on the disk.

The Ernst potential for the dust disk solution is given by the expression
\begin{equation}
f = \exp
\left\{\mu\left[\,\int\limits_{X_1}^{X_a}\frac{X^2\,dX}{W} +  
\int\limits_{X_2}^{X_b}\frac{X^2\,dX}{W} - 
\int\limits_{-\rm i}^{\rm i}\frac{hX^2\,dX}{W_1}\right]\right\},
\label{f}
\end{equation}
with
\begin{equation}
W = W_1W_2, \quad W_1=\sqrt{(X-\zeta/\rho_0)^2+(\rho/\rho_0)^2},\quad
W_2=\sqrt{1+\mu^2(1+X^2)^2},
\end{equation}
\begin{equation}
h = \frac{\ln\left(\sqrt{1+\mu^2(1+X^2)^2}+\mu(1+X^2)\right)}
{\pi {\rm i}\, \sqrt{1+\mu^2(1+X^2)^2}},
\end{equation}
\begin{equation}
X_1^2 = \frac{\rm i - \mu}{\mu}, \quad  X_2^2 = -\frac{\rm i + \mu}{\mu}
\quad (\Re X_1 < 0, \quad \Re X_2 > 0).
\end{equation}
The upper integration
limits $X_a$ and $X_b$ in the first two integrals are hyperelliptic
functions of two variables $u$ and $v$ (i.~e. `ultraelliptic' functions)
defined by 
\begin{equation}
\int\limits_{X_1}^{X_a}\frac{dX}{W} + \int\limits_{X_2}^{X_b}\frac{dX}{W} = u,
\quad \int\limits_{X_1}^{X_a}\frac{X\,dX}{W} + 
\int\limits_{X_2}^{X_b}\frac{X\,dX}{W} = v,
\label{jacobi}
\end{equation}
where the variables $u$ and $v$ are given by
\begin{equation}
u = \int\limits_{-\rm i}^{\rm i}\frac{h\,dX}{W_1}, \quad
v = \int\limits_{-\rm i}^{\rm i}\frac{hX\,dX}{W_1}.
\label{uv}
\end{equation}
The third integral in (\ref{f}) as well as the integrals in (\ref{uv}) have
to be taken along the imaginary 
axis in the complex X--plane with $W_1$ and $h$
fixed according to $\Re W_1 <0$ (for $\rho$, $\zeta$ outside the disk)
and $\Re h =0$.

Eqs.~(\ref{jacobi}) have exactly the form of Jacobi's famous
inversion problem 
\cite{J}, which was 
solved by G\"opel \cite{G} and Rosenhain \cite{R}
who were able to express $X_a(u,v)$ and $X_b(u,v)$ in terms of (ultraelliptic)
theta functions. In our case the moduli of the theta functions depend on
$\rho/\rho_0$, $\zeta/\rho_0$, and $\mu$.

It should be noted that the integrations from $X_1$ to $X_a$ resp.~$X_2$ to
$X_b$ in (\ref{f}) and (\ref{jacobi}) have to be performed 
{\it along the same curve} 
in the two--sheeted Riemann surface associated with $W(X)$. This leads to
a unique Ernst potential $f(\rho/\rho_0,\zeta/\rho_0,\mu)$.
More technical details and final expressions for the metric in terms
of theta functions will be presented
in a subsequent paper. 
\paragraph{The Newtonian limit $\mu \ll 1$.}
At $\mu=0$, the solution may be expanded into a power series in $\mu^{1/2}$:
\begin{equation}
f = 1+\sum\limits_{n=1}^{\infty} f_n\mu^{(n+1)/2}. 
\end{equation}
The coefficients $f_n(\rho/\rho_0, \zeta/\rho_0)$  turn out to
be elementary
functions. A particularly simple result is obtained
for $f_1$ to $f_4$. From Eqs.~(\ref{f}) to (\ref{uv}) we get
\begin{equation}
X_a-X_1 = {\cal O}(\mu^{3/2}),\quad X_b-X_2 = {\cal O}(\mu^{3/2}),
\end{equation}
and
\begin{equation}
f = \exp\left\{-\mu\int\limits_{-\rm i}^{\rm i}\frac{h(X-X_1)(X-X_2)\,dX}{W_1}
+{\cal O}(\mu^3)\right\}.
\end{equation}
Expanding $h$, $X_1$ and $X_2$ one can easily determine $f_1$ to $f_4$.
(Note that $X={\cal O}(1)$ in this integral.) The result for $f_1$ is
\begin{equation}
f_1 = \frac{1}{\pi{\rm i}}\int\limits_{-\rm i}^{\rm i}\frac{(1+X^2)\,dX}
{W_1} =  -\frac{1}{\pi}\,\left\{\frac{4}{3}\,\cot^{-1}\xi+
\left[\xi-(\xi^2+\frac{1}{3})\cot^{-1}\xi\right](1-3\eta^2)\right\}
\end{equation}
with elliptic coordinates $\xi$, $\eta$ defined by
\begin{equation}
\rho=\rho_0\,\sqrt{1+\xi^2}\sqrt{1-\eta^2},\quad
\zeta=\rho_0\xi\eta,\quad 0\le\xi<\infty,\quad -1\le\eta\le 1.
\end{equation}
Now, the Newtonian limit $U_{Ml}$ can be read off from the first order
relation
\begin{equation}
g_{tt} = - \Re f = -(1 + \mu f_1) = -(1 + \frac{2U_{Ml}}{c^2}).
\end{equation}
Obviously, $U_{Ml} = c^2 \mu f_1/2$ is the Maclaurin solution,
i.~e.~the Newtonian solution of
the problem of a rigidly rotating disk of dust. Note that in this limit 
\begin{equation}
\mu = -\frac{2U_{Ml}(\rho=0,\zeta=0)}{c^2} = \frac{2\Omega^2\rho_0^2}{c^2}.
\end{equation}
(For the sake of clarity we have reintroduced the velocity of light $c$
in the last two equations.)
 
\paragraph{Discussion.}
In the parameter range $0<\mu<\mu_0=4.62966184\dots$
\cite{NM2} the solution is asymptotically flat and regular everywhere
outside the disk. Its baryonic surface mass density is positive and
vanishes at the rim of the disk. For $\mu\to\mu_0$ ($\Omega$ finite)
and $\rho^2+\zeta^2\ne 0$ it approaches exactly the extreme
Kerr solution (mass $M=1/\,2\Omega$). 
The meaning of the solution for $\mu>\mu_0$
deserves further discussion.

The analytic expressions above allow for a detailed investigation of
characteristic properties, as for instance the formation of toroidal
ergospheres, dragging effects, motion of test particles, and stability.

\end{document}